\documentclass[twocolumn,,amssymb,amsmath,aps,showpacs,floatfix,nofootinbib,showpacs,balancelastpage]{revtex4}
\usepackage{amsmath}
\usepackage{amssymb}

\usepackage{hangcaption,fancybox,feynmp}
\usepackage{indentfirst}
\usepackage{graphicx}
\usepackage{epsfig,subfigure,psfrag}
\usepackage{mathrsfs}

\begin{document}


\title{New Color-Octet Vector Boson?}
\author{Bo Xiao$^{1}$\footnote{E-mail:homenature@pku.edu.cn},
You-kai Wang$^{1}$\footnote{E-mail:wangyk@pku.edu.cn},
 and Shou-hua
Zhu$^{1,2}$\footnote{E-mail:shzhu@pku.edu.cn} }

\affiliation{ $ ^1$ Institute of Theoretical Physics $\&$ State Key
Laboratory of Nuclear Physics and Technology, Peking University,
Beijing 100871, China \\
$ ^2$ Center for High Energy Physics, Peking University, Beijing
100871, China }

\date{\today}

\begin{abstract}

Both CDF and D0 at Tevatron reported the measurements of
forward-backward asymmetry in top pair production, which showed
possible deviation from the standard model QCD prediction. In this
paper, we show that a new color-octet massive vector boson with mass
just above twice that of top quark can simultaneously account for
the asymmetry and differential distribution $d\sigma/dM_{t\bar t}$
in top pair production, without conflict with other measurements for
example di-jet production. The new particle can be discovered and
studied at the more powerful Large Hadron Collider.

\end{abstract}


\maketitle

Discovering new particle is one of the most important goals for
higher and higher energy colliders, such as Tevatron at Fermilab and
Large Hadron Collider (LHC) at CERN. Top quark was discovered in
1995 at Tevatron. Since the important discovery,  measuring
properties of top quark is one of the most active field in high
energy physics. The endeavor is justified because the top quark is
the heaviest ever known fermion and is thought to be related to
mechanism of electro-weak symmetry breaking and physics beyond the
standard model (SM).  Most of measured properties such as mass,
width, production rate are consistent with SM predictions, however
the CDF and D0 Collaboration have observed possible deviation on
forward-backward (FB) asymmetry in top quark pair production.  In
this paper we will show that the deviation can be due to the new
color-octet massive vector boson.

At $t\bar t$ frame the FB asymmetry in top quark pair production $A_{FB}$ is
defined as \begin{equation}
\begin{array}{rl}
A_{FB}&=\frac{\sigma( \Delta Y>0)-\sigma(\Delta Y<0)}{\sigma(\Delta
Y>0)+\sigma(\Delta Y<0)} \equiv \frac{\sigma_A}{\sigma}, \label{originAFB}
\end{array}
\end{equation}
where $\Delta Y\equiv Y_t-Y_{\bar t}$ is the difference between
rapidity of the top and anti-top quark, which is invariant under
$t\bar t$ or $p\bar{p}$ rest frame.

The measurements of CDF and D0 are \cite{CDFAfb5.3,ICEHP2010Shary},
\begin{equation} \begin{array}{rl}
A_{FB}^{CDF} &= 0.158\pm 0.072\pm 0.017, {\rm with}  \ 5.3 fb^{-1};  \\
A_{FB}^{D0}  &= 0.08\pm 0.04\pm 0.01, {\rm with}  \ 4.3 fb^{-1}.
\end{array}
\end{equation}
 The measurements are consistent with
previous ones \cite{Abazov:2008,Aaltonen:2008hc,Aaltonen:2009iz},
but reveal about a $2\sigma$ deviation from the SM's prediction
\cite{Kuhn:1998PRL, Kuhn:1998PRD, Kuhn:2008, G.Sterman:2008,
Z.G.Si:2010, L.L.Yang:2010}. The discrepancy has inspired lots of
new physics discussions \cite{Frampton:2009rk, Shu:2009xf,
Jung:2009jz, Cheung:2009ch, Cao:2010zb, Djouadi:2009nb, Jung:2009pi,
Cao:2010, Barger:2010, Arhrib:2009hu}. These discussions can be
roughly classified into two categories. One category is by
introducing a $Z'$ or $W'$ which have flavor changing couplings with
fermions. The flavor changing coupling among $Z'$, top and up quarks
can induce a t-channel diagram of $u\bar u\rightarrow t\bar t$ which
contributes to $A_{FB}$ via the interference with usual QCD tree
diagrams. The other category is by introducing a
heavy($>1\text{TeV}$) axial-gluon. It induces a s-channel diagram of
$q\bar q\rightarrow t\bar t$ which contributes to $A_{FB}$ via the
interference with usual QCD tree diagrams and/or itself.  However,
for the new physics in the first category, $d\sigma/d M_{t\bar t}$
distribution for top pair production is violated greatly
\cite{Jung:2009jz}, even after including more higher-order effects
\cite{Xiao:2010hm}. Other severe constraint comes from the
measurement of same-sign top production rate at Tevatron
\cite{Aaltonen:2008hx,Jung:2009jz}. For the second category of new
physics, the suitable parameters to account for all existing
measurements can hardly be found \cite{Frampton:2009rk,
Chivukula:2010fk}, especially to satisfy the constraint from the
high $M_{t\bar t}$ region. The lesson from these investigations
\cite{Frampton:2009rk, Shu:2009xf, Jung:2009jz, Cheung:2009ch,
Cao:2010zb, Djouadi:2009nb, Jung:2009pi, Cao:2010, Barger:2010,
Arhrib:2009hu} is that it is very difficult to account for $A_{FB}$
without distorting the shape of  $d\sigma/d M_{t\bar t}$. Totally
new idea is indispensable. How about to introduce {\em s-channel}
diagram induced by a {\em low mass} color-octet vector boson? In
this case, the $A_{FB}$ can be induced by interference with the
corresponding QCD diagrams, at the same time the shape of $d\sigma/d
M_{t\bar t}$ for high $M_{t\bar t}$ is minimally affected. In this
paper we will show that  by introducing a new color-octet massive
vector boson (denoted by $Z_C$ hereafter) with the mass $M_C$ just
above $2 m_t$, the measured $A_{FB}$ and $d\sigma/d M_{t\bar t}$ can
be accounted for. Furthermore such kind of new particle are
compatible with all other measurements.

\begin{figure}[htbp]

\begin{center}
\includegraphics[width=6cm]
{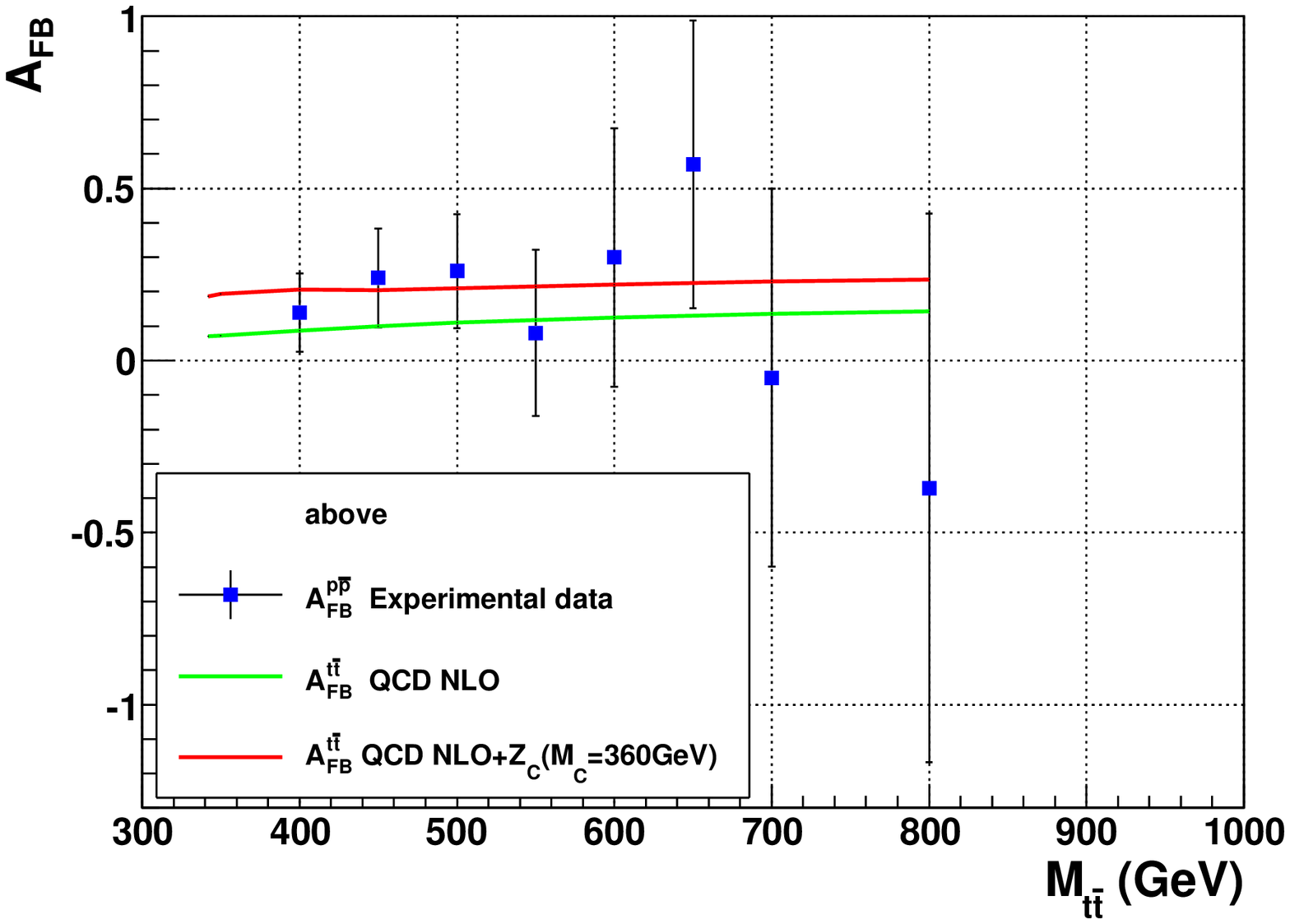}
\includegraphics[width=6cm]
{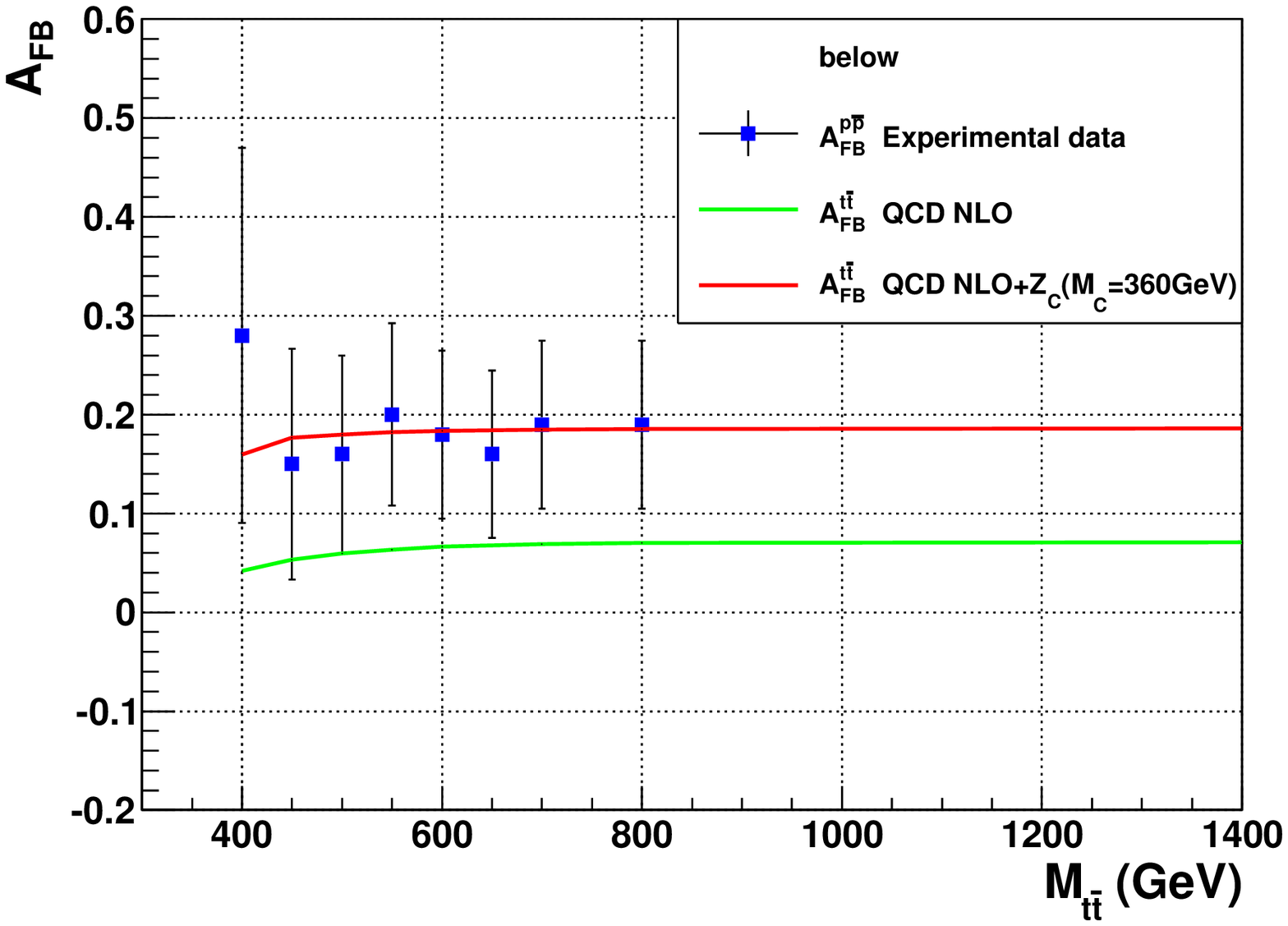}
\end{center}
\begin{center}
\includegraphics[width=6cm]
{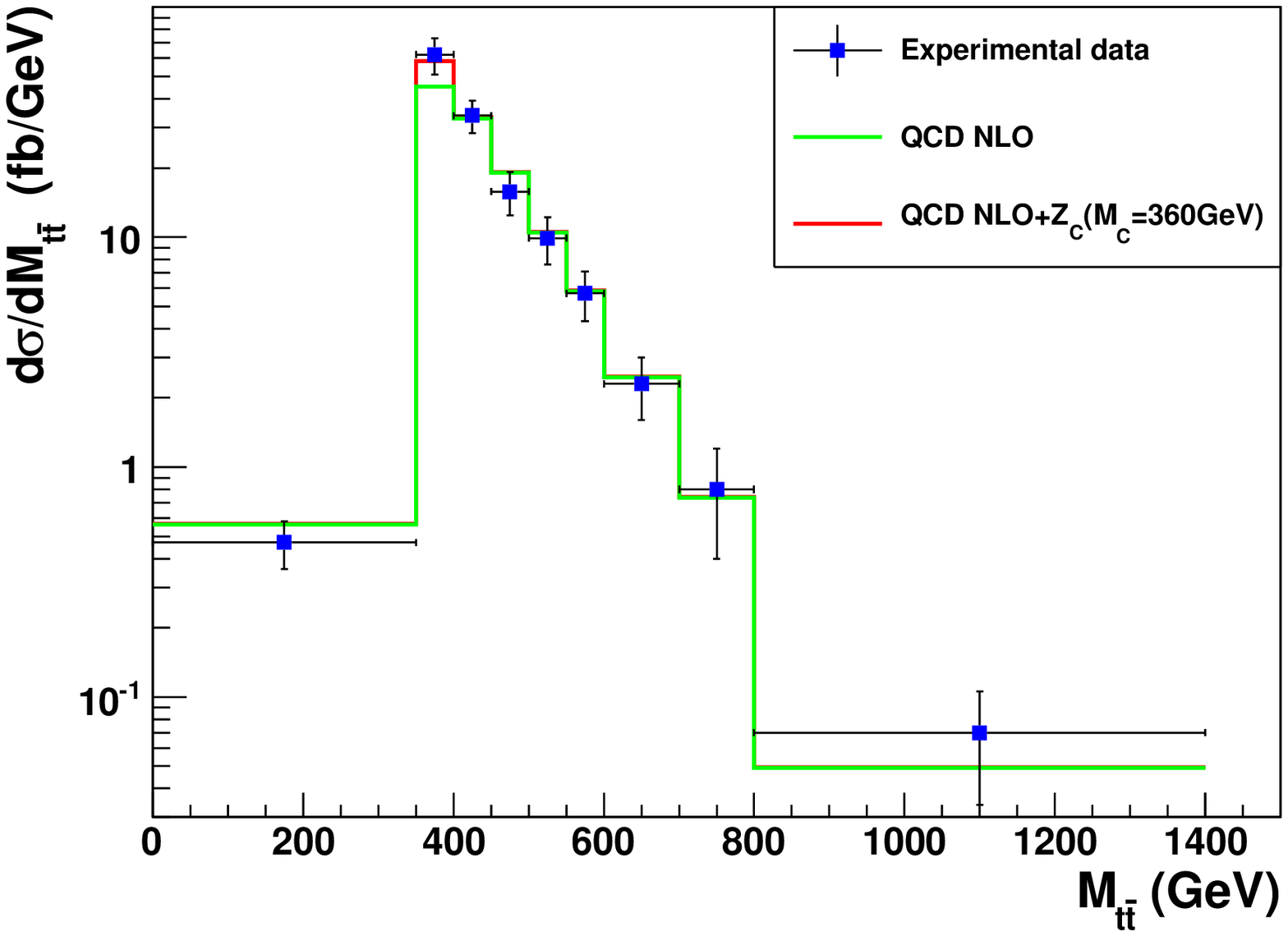}
\end{center}

\caption{\label{fig1} $A_{FB}$ and $d\sigma/dM_{t\bar t}$ as a
function of $ M_{t\bar t}$ in the SM and $Z_C$+SM. For the new
particle $Z_C$, $M_C=360 \text{GeV}$ is chosen. Other allowed $M_C$
induces the similar behavior. Experimental data is also shown. Here
`above' and `below' represent that $A_{FB}$ is calculated above or
below a specific $ M_{t\bar t}$ and  $m_t= 170.9$ GeV. }

\end{figure}

In Fig. \ref{fig1}, the measurements \cite{CDFbelowabove} and
predictions in the SM for $A_{FB}$ and $d\sigma/d M_{t\bar t}$ are
depicted. From the figures it is quite natural to expect that extra
contributions to {\em both} $A_{FB}$ and $d\sigma/d M_{t\bar t}$ are
in the low $M_{t\bar t}$ region. Adopting the central values, the
extra asymmetric cross section $\delta \sigma_A$ of $\sim 800$ fb is
required. At the same time additional $\sim 700$ fb cross section
$\delta \sigma$ is also required in the 350-400 GeV bin, which is
hardly accommodated by threshold resummation \cite{G.Sterman:2008,
L.L.Yang:2010}. The resummation effects will increase the
distribution evenly over the whole $M_{t\bar t}$ region.

Color-octet massive particle conjectured in physics beyond the SM
(BSM) is nothing new. For example in supersymmetric models there is
gluino, namely the supersymmetric partner of gluon. Gluinos couple
with the color particle in the strength of strong interaction which
is well described by QCD. Another example is the above-mentioned
axial-gluon which is the mediator for the new gauge group. Such
axigluon has been proposed to account for the FB asymmetry in top
pair production with the mass of axial-gluon at $\mathcal {O}(1)$
TeV. However imposing the constraints from differential distribution
of top pair invariant mass, di-jet production measurement, as well
as other low energy measurements, such proposals seem to be
disfavored. The failure to account for FB asymmetry in top pair
production utilizing axial-gluon lies in the implicit assumption of
the couplings with top quark and light quarks, which are at
$\mathcal {O}(g_S)$ i.e. the strong coupling constant.

In the phenomenological model we introduce  a new color-octet axial
massive vector boson $Z_C$. The coupling with the top-quark is taken
to be $-i{g_t}{\gamma ^\mu }{\gamma ^5}{T^a}$, and coupling with
other quarks to be $-i{g_q}{\gamma ^\mu }{\gamma ^5}{T^a}$
\footnote{In the realistic model, bottom quark is usually grouped
with top quark. The assumption here does not change our main
results. }.

\begin{figure}[htbp]

\begin{center}
\includegraphics[width=3.5cm]
{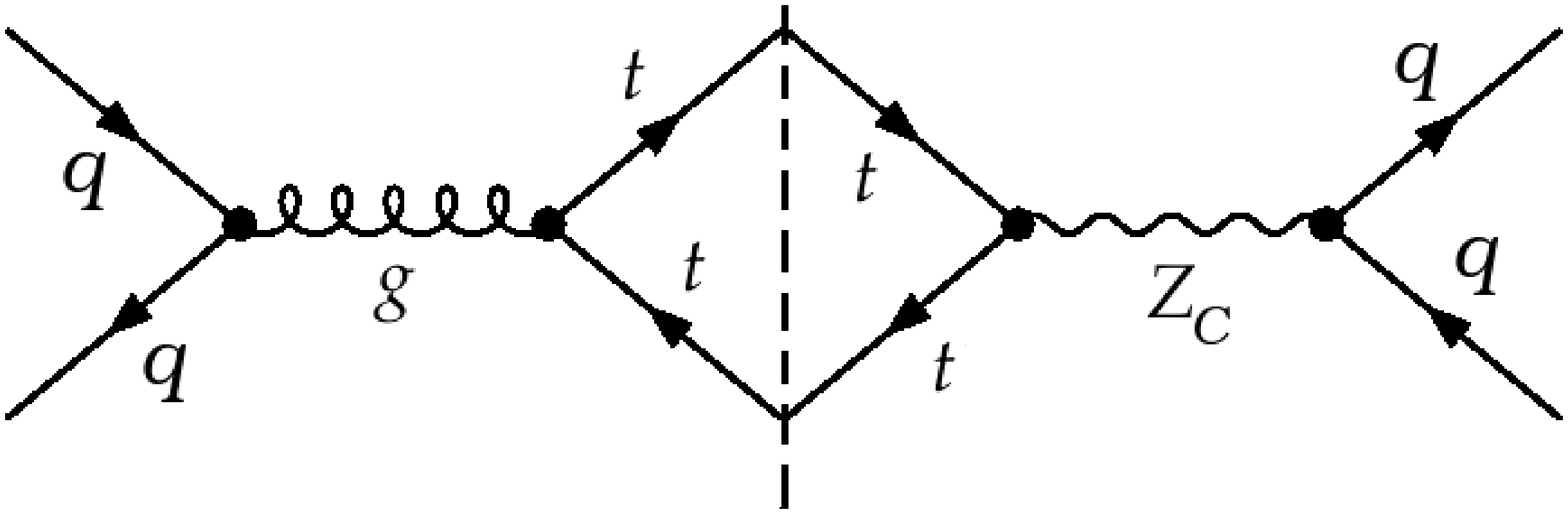}
\includegraphics[width=3.5cm]
{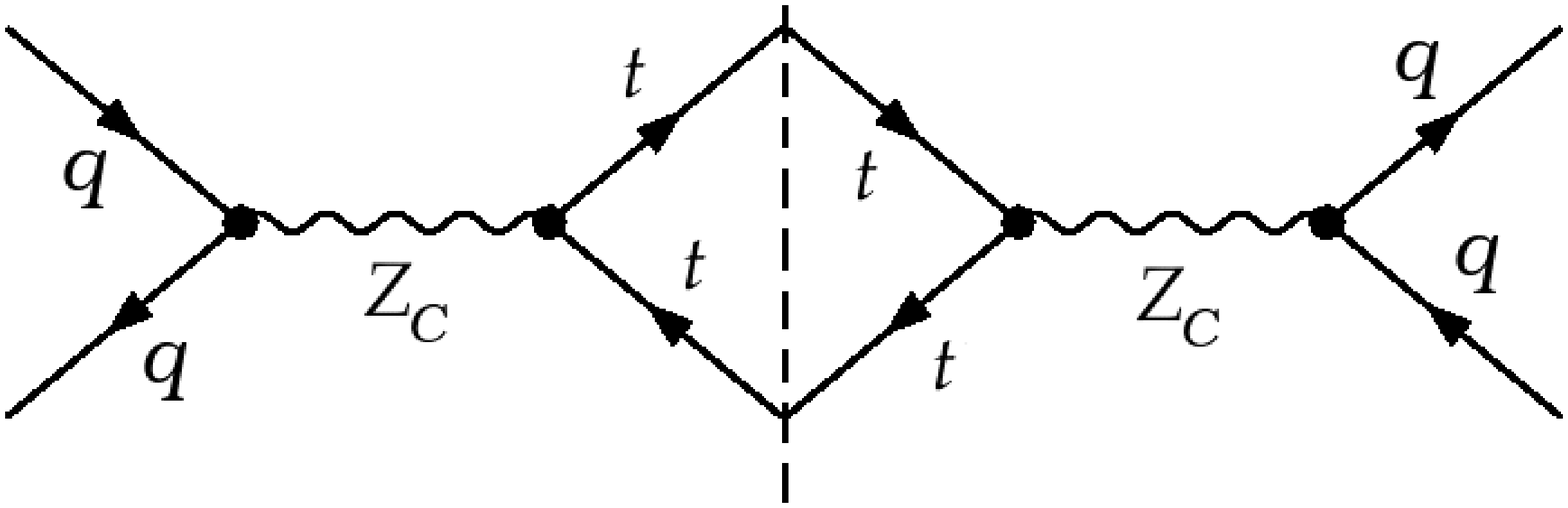}
\end{center}

\caption{\label{fig2} Left diagram: interference between usual QCD
tree diagram and $Z_C$ induced tree diagram. Right diagram: square
of $Z_C$ induced tree diagram.}

\end{figure}

In this model, $\delta \sigma_A$ arises from the left diagram and
$\delta \sigma$ the right one in Fig. \ref{fig2}. The analytical
expression of $\sum_{\text{Color,Spin}}{|M|^2}$ for the left diagram
can be written as
\begin{equation}
32\pi
C_AC_F\alpha_sg_qg_t\frac{(s-M_C^2)s}{(s-M_C^2)^2+\Gamma_C^2M_C^2}\cos\theta
\beta, \label{acs}
\end{equation}
where $C_A=3$, $C_F=4/3$, $\beta\equiv \sqrt{1-4m_{t}/s}$ and
$\Gamma_C$ is the total width of $Z_C$
\[
\Gamma_C=\sum_i^{2m_i<M_C}{\frac{g_i^2}{4\pi}\frac{C_F}{8}(1-4m_i^2/M_C^2)^\frac{3}{2}M_C}.
\]
$\sum_{\text{Color,Spin}}{|M|^2}$ for the right diagram  can be
written as
\begin{equation}
2C_AC_F(g_qg_t)^2\frac{s^2}{(s-M_C^2)^2+\Gamma_C^2M_C^2}(1+\cos^2\theta)\beta^2. \label{scs}
\end{equation}

In the following we will determine the model parameters namely $M_C$
and the two couplings constants $g_t$ and $g_q$, utilizing the
$\delta\sigma_A=800$ fb and $\delta \sigma=700$ fb. The procedure is
divided into two steps.

Step one: Determining the parameters using  $\delta\sigma_A$ and
$\delta\sigma$. There are three additional free parameters $M_C$,
$g_t$, $g_q$.  For the convenience of numerical analysis, we first
transform the three free parameters into $M_C$, $g_t g_q$ and
$g_q/g_t$. All possible $g_q/g_t$, $M_C$ and $g_tg_q$ can be found
by scanning the parameter space.

Step two: Limiting the parameters $g_q/g_t$, $M_C$ and $g_t g_q$ by
requiring that the extra contribution from $Z_C$ can improve the
agreement between theoretical predictions and experimental
measurements, namely the distributions of $d\sigma/dM_{t\bar t}$ and
$A_{FB}$. The possible parameter $g_q/g_t$ is confined to a very
narrow region $0.0040 \leq g_q/g_t \leq 0.0044$. At the same time,
$M_C$ and $g_t g_q$ are approximately written as
$$
g_q g_t \simeq  0.2 \frac{M_C-290 [GeV]}{m_t}; 350~\text{GeV} \leq M_C \leq 380~\text{GeV}.
$$
In Fig. \ref{fig1} we show the excellent agreement between
theoretical predictions and data after including extra contributions
from $Z_C$  with the possible parameter set $M_C= 360$ GeV,
$g_q/g_t=0.0042$ and $g_q g_t = 0.082$. As for $g_q$, extra
constraint may arise from the measurements of di-jet production
\cite{Aaltonen:2008dn}. However it is obviously that the required
$g_q$ here is much less than the limit.

We should emphasize that the numerical results here is just for the
illustration purpose. The generic features for other allowed $\delta
\sigma_A$ and $\delta \sigma$ due to the uncertainties of
measurements are the same, namely (1) color octet vector boson with
mass just above $2m_t$ can improve the agreement between the
predictions and data; (2) the coupling among new vector boson with
top is much larger than that of light quarks; (3) the couplings of
$Z_C$ should be axial-vector like. The third point can be understood
as following. If taking the more generic coupling of $Z_C$ as
$({g_V}{\gamma ^\mu } + {g_A}{\gamma ^\mu }{\gamma ^5})$ in the
beginning, we find that $g_V$ has to be much smaller than $g_A$ in
order to account for $\delta \sigma$ and $\delta \sigma_A$
simultaneously.  The underlying reason is very simple. The $\sum
|M|^2$ of $\delta \sigma_A$ brought by $g_A$ is proportional to
$\beta$ (cf. Eq. \ref{acs}), while the $\sum |M|^2$ of $\delta
\sigma$ brought by $g_V$ doesn't have this feature. For $M_{t\bar
t}$ in 350-400 GeV, $\beta$ is small, therefor $g_V$ has to be much
smaller than $g_A$ in order not to bring much more $\delta \sigma$
than $\delta \sigma_A$. It is the observed $\delta \sigma$ and
$\delta \sigma_A$ that fix the coupling $g_V$ much less than $g_A$.

The key difference between the proposed model and the axial-gluon
model introduced in Ref. \cite{Frampton:2009rk} is that, for the
latter the mass of axial-gluon is quite heavy ($ \mathscr{O}(1)
\text{TeV}$) so the the axial couplings with top and other quarks
have the {\em opposite} sign in order to induce the
 positive $A_{FB}$ from interference,
while in this model, the $M_C$ is assumed just above
$2 m_{t}$ and the axial-vector couplings with top-quark and the
other quarks have the {\em same } sign.

\begin{figure}[htbp]

\begin{center}
\includegraphics[width=5cm]
{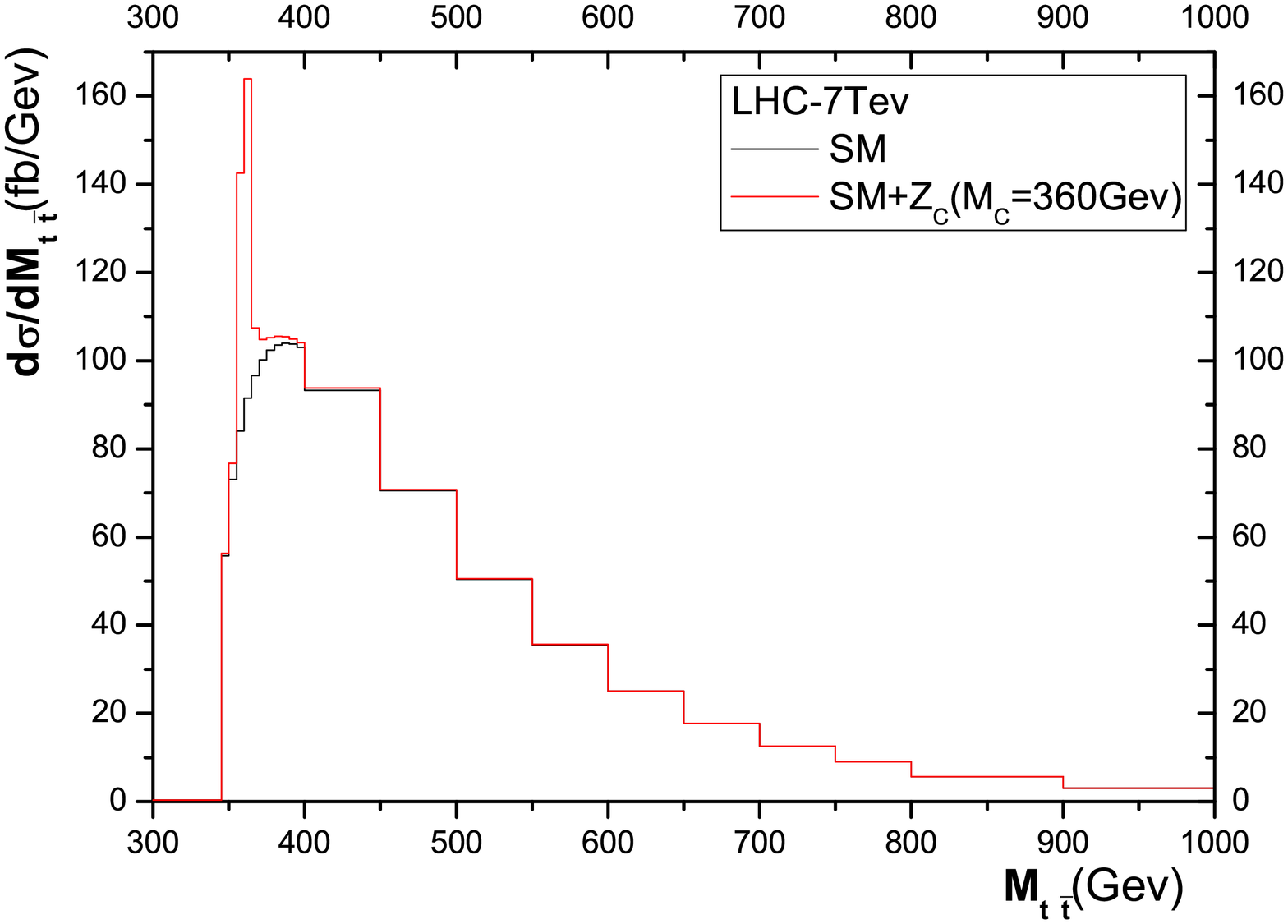}
\end{center}
\begin{center}
\includegraphics[width=5cm]
{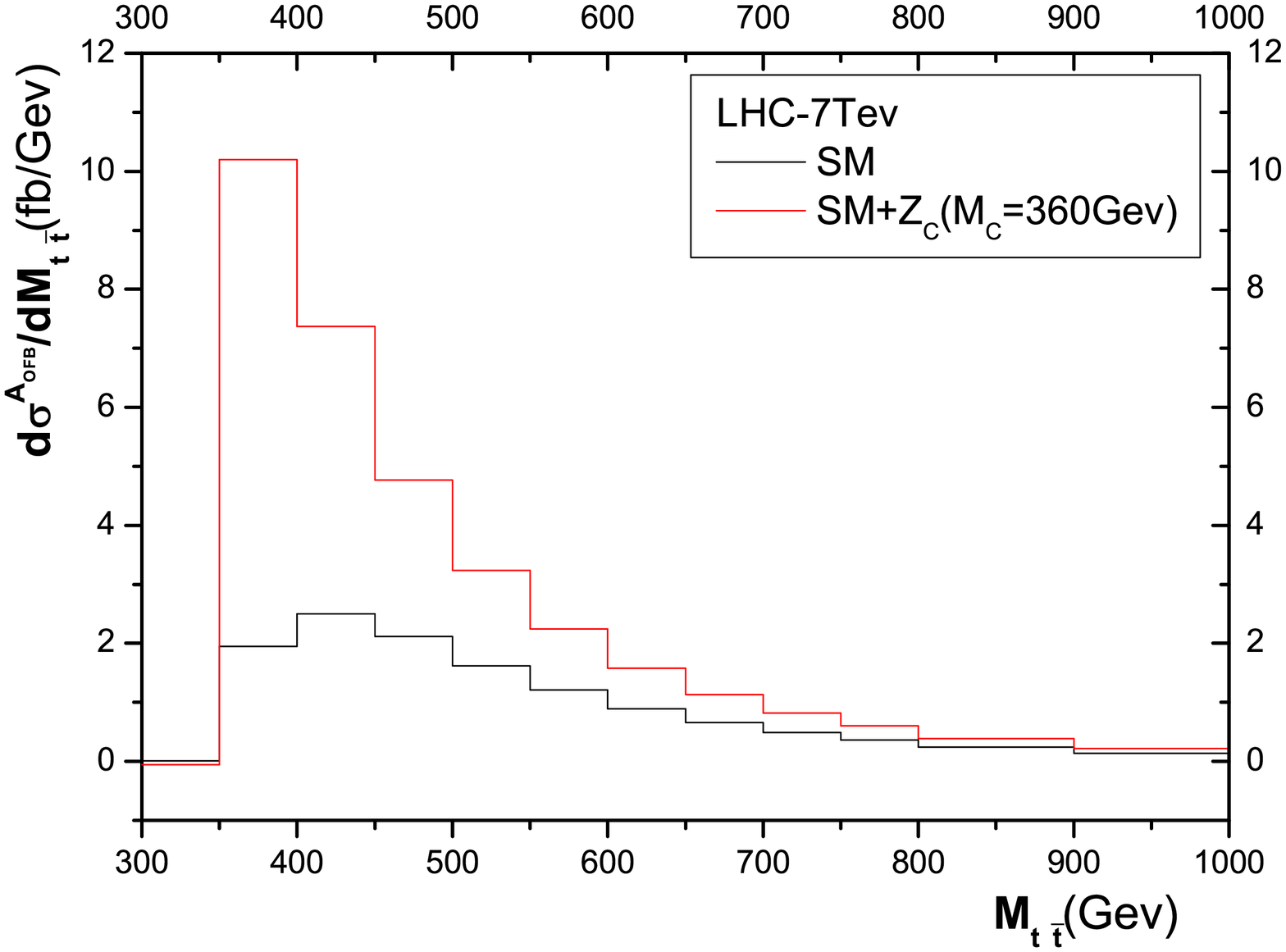}
\end{center}

\caption{\label{fig4} Differential cross section $d\sigma/dM_{t\bar
t}$ (up figure) and differential one-side asymmetric cross section
$d\sigma^{A_{\text{OFB}}}/dM_{t\bar t}$ (bottom figure) in
$Z_C$+SM and SM. Here $M_C=360\text{GeV}$ is chosen, and the other allowed $M_C$ give the similar
behavior. Note that $P_{t\bar t,\text{cut}}^z$, namely the top pair momentum cut in z-direction, is used for  $\sigma$ and $\sigma^{A_{\text{OFB}}}$ calculations.}

\end{figure}

Tevatron has shown sign of the new color-octet vector boson, and it
is quite natural to explore how to discover and study such kind of
new particle at the more powerful LHC. In Fig. \ref{fig4}, we show
the differential cross section $d\sigma/dM_{t\bar t}$ as a function
of $M_{t\bar t}$ in $Z_C$ model. It is clear that the top-antitop
production cross section is larger than that of in the SM,
especially in the low $M_{t\bar t}$ region via the sub-process
$q\bar q\rightarrow t\bar t$ at LHC. Due to the Yang theorem,
on-shell $Z_C$ does not contribute to top pair production in
gluon-gluon fusion, which is the main production mechanism in the
SM. The bump of $Z_C$ at $M_{t\bar t}$ should appear after
collecting enough data samples.

The coupling properties of new particle can be studied via the
angular distributions. However the usual forward-backward asymmetry
defined at Tevatron is not applicable at LHC. The reason is that LHC
is the proton-proton collider, there is no preferred direction which
is contrary to the proton anti-proton colldier Tevaton. Fortunately,
there are some solutions \cite{Kuhn:1998PRL, Kuhn:1998PRD,
Kuhn:2008, Rodrigo:2008, Wang:2010du}. Based on Ref.
\cite{Wang:2010du} we will study the one-side forward-backward
asymmetry in the $Z_C$ model. The differential one-side asymmetric
cross-section $d\sigma^{A_{\text{OFB}}}/dM_{t\bar t}$ is shown in
Fig. \ref{fig4}, which is important to distinguish $Z_C$ model from
other possible contributions. The integrated one-side asymmetric
cross sections $\sigma^{A_{\text{OFB}}}$ are depicted in Tab. 1.
Also shown is the signal significance $Sig$ which is defined as
$$
Sig=\frac{\sigma^{A_{\text{OFB}}}_{\text{SM}+Z_C}-\sigma^{A_{\text{OFB}}}_{\text{SM}}}{\sqrt{\sigma_{\text{SM}}}}\sqrt{
\mathscr{L}},
$$
where $\mathscr{L}$ is taken to be $10\text{fb}^{-1}$.
In order to beat the huge QCD backgrounds, especially the ones from
gluon-gluon fusion, selection cuts are necessary both for $\sigma$
and for $\sigma^{A_{\text{OFB}}}$\cite{Wang:2010du}. The optimal
choice of cuts is $P_{t\bar t,\text{cut}}^z=600\text{GeV}$ for LHC
at 7 TeV and $P_{t\bar t,\text{cut}}^z=1.2\text{TeV}$ for LHC at 14
TeV \cite{Wang:2010du}. From the figure and table, LHC can discover and measure the coupling nature of such kind of $Z_C$ with quite low integrated luminosity.

\begin{table}[htb]

\caption{\label{Afb} Total one-side asymmetric cross section (fb) in
$Z_C$+SM. In the SM, $\sigma_{\text{SM(LO)}} \simeq $  $22.2\times
10^3$, $75\times 10^3$fb and $\sigma_{\text{SM}}^{A_{\text{OFB}}}
\simeq$ 650, 1650fb for $\sqrt{s}=7$ and $14$ TeV respectively. Note
that $P_{t\bar t,\text{cut}}^z$ is used for all the $\sigma$ and
$\sigma^{A_{\text{OFB}}}$ calculations.}

\center \small{

\begin{tabular}{ccc}
\hline\hline
 &
\begin{tabular}{c}
$\sigma_{\text{SM}+Z_C}^{A_{\text{OFB}}}(M_C=)$\\
\begin{tabular}{ccc}
355.0&360.0&370.0
\end{tabular}\\
\end{tabular} &
\begin{tabular}{c}
$\text{$Sig$}(M_C=)$\\
\begin{tabular}{ccc}
355.0&360.0&370.0
\end{tabular}\\
\end{tabular}\\
\hline 7 TeV&
\begin{tabular}{ccc}
1623  &1670 &1716  \\
\end{tabular} &
\begin{tabular}{ccc}
20.63 &21.62 &22.61 \\
\end{tabular}\\
\hline 14 TeV&
\begin{tabular}{cccc}
3971 &4096 &4245 \\
\end{tabular} &
\begin{tabular}{ccc}
26.69 &28.13 &29.85 \\
\end{tabular}\\
\hline\hline
\end{tabular}

}

\end{table}

To summarize, both  CDF and D0 at Tevatron reported the measurements
of forward-backward asymmetry in top pair production. Theoretically
such asymmetry is due to the higher-order QCD processes in the SM.
The measurements showed possible deviation from the theoretical
prediction. In this paper a phenomenological model which contains
the new color-octet massive vector boson $Z_C$ is proposed. When the
mass of $Z_C$ is just above twice that of top quark and the
couplings are appropriately chosen, the asymmetry and distribution
of $M_{t\bar t}$ in top pair production can be explained
simultaneously, without conflict with other measurements for example
di-jet production.

We would like to emphasize the implications of our study for
model-building. The requirements for the new massive color-octet
vector boson $Z_C$ are (1) $M_C$ is just above $2 m_t$; (2) the
nature of couplings among $Z_C$ and quarks is axial-vector like; (3)
the axial coupling of $Z_C$ with top quark is much larger than that
with light quarks, but are of the same sign, which is contrary to
the conventional axial-gluon models. These features indicate that
$Z_C$ can be intimately correlated with conjectured top quark pair
condensate, and even the mechanism of electro-weak symmetry
breaking. We are not aware of any models in literature which have
such features. Hopefully the Tevatron asymmetry measurements are the
sign for the new particle and true underlying mechanism will be
uncovered at the LHC.

{\em Acknowledgements:} This work was supported in part by the
Natural Sciences Foundation of China (Nos. 10775001, 10635030 and
11075003).

\end{document}